\documentclass[final,5p,times,twocolumn]{elsarticle}

\usepackage[T1]{fontenc}
\usepackage[utf8]{inputenc}

\usepackage{amsmath}
\usepackage{amsfonts}
\usepackage{amssymb}
\usepackage{amsxtra}
\usepackage{array}
\usepackage{color}
\usepackage{dcolumn}
\usepackage{graphicx}
\usepackage{hepunits}
\usepackage{subcaption}
\usepackage{xspace}
\usepackage{CJKutf8}

\definecolor{purple}{rgb}{0.5,0,0.5}
\definecolor{blue}{rgb}{0.0,0,0.9}
\definecolor{prdblue}{rgb}{0.133,0.118,0.498}
\usepackage[colorlinks=true, pdfstartview=FitV, linkcolor=prdblue, citecolor= prdblue, urlcolor=prdblue]{hyperref}

\usepackage[mathscr,scaled=1.15]{urwchancal}
\DeclareFontFamily{OT1}{pzc}{}
\DeclareFontShape{OT1}{pzc}{m}{it}%
{<-> s * [1.15] pzcmi7t}{}
\DeclareMathAlphabet{\mathpzc}{OT1}{pzc}{m}{it}

\biboptions{sort&compress}

\journal{Journal of Subatomic Particles and Cosmology}

\begin{document}
\begin{CJK*}{UTF8}{gbsn}
\begin{frontmatter}

\title{$\,$\\[-6ex]\hspace*{\fill}{\normalsize{\sf\emph{Preprint no}.\
NJU-INP 115/26}}\\[1ex]
Insights into Meson and Baryon Structure using Continuum Schwinger Function Methods}

\author[ECT]{Daniele Binosi%
    $\,^{\href{https://orcid.org/0000-0003-1742-4689}{\textcolor[rgb]{0.00,1.00,0.00}{\sf ID}}}$}

\author[NJU,INP]{Craig D.\ Roberts%
       $^{\href{https://orcid.org/0000-0002-2937-1361}{\textcolor[rgb]{0.00,1.00,0.00}{\sf ID}},}$$^\ast$}

\author[HZDR]{Zhao-Qian Yao (姚照千)%
       $\,^{\href{https://orcid.org/0000-0002-9621-6994}{\textcolor[rgb]{0.00,1.00,0.00}{\sf ID}}}$}

\address[ECT]{European Centre for Theoretical Studies in Nuclear Physics
            and Related Areas  (\href{https://ror.org/01gzye136}{ECT*}), Villa Tambosi, Strada delle Tabarelle 286, I-38123 Villazzano (TN), Italy}
\address[NJU]{
School of Physics, \href{https://ror.org/01rxvg760}{Nanjing University}, Nanjing, Jiangsu 210093, China}
\address[INP]{
Institute for Nonperturbative Physics, \href{https://ror.org/01rxvg760}{Nanjing University}, Nanjing, Jiangsu 210093, China}

\address[HZDR]{\href{https://ror.org/01zy2cs03}{Helmholtz-Zentrum Dresden-Rossendorf}, Bautzner Landstra{\ss}e 400, D-01328 Dresden, Germany

%
\href{mailto:cdroberts@nju.edu.cn}{cdroberts@nju.edu.cn} 
\\[1ex]
Date: 2026 January 11\\[-5ex]
}

\cortext[1]{Speaker and corresponding author}

\begin{abstract}
The bulk of visible mass is supposed to emerge from nonperturbative dynamics within quantum chromodynamics (QCD).  Following years of development and refinement, continuum and lattice Schwinger function methods have recently joined in revealing the three pillars that support this emergent hadron mass (EHM); namely, a nonzero gluon mass-scale, a process-independent effective charge, and dressed-quarks with running masses that take constituent-like values at infrared momenta. One may argue that EHM and confinement are inextricably linked; and theory is now working to expose their manifold expressions in hadron observables and highlight the types of measurements that can be made in order to validate the paradigm. This contribution sketches these ideas via the unified explanation of pion and proton electromagnetic and gravitational form factors.
\end{abstract}


\end{frontmatter}
\end{CJK*}


\section{Introduction}
Internationally, high-energy nuclear and particle physics communities are engaged in determining whether quantum chromodynamics (QCD) is the theory that describes strong interactions in the Standard Model (SM).
Perturbatively, QCD works.
The question remains, however, is that is the end of the story or is QCD also nonperturbatively well-defined so that it describes all of strong interaction physics?
As distance markers on the road to an answer, the communities have identified solutions to the following three of Nature's most fundamental puzzles:
(\emph{a}) What is the origin of $m_N$, the proton mass, which sets the basic mass unit for all visible matter and, whatever that is, why is the pion, with its small lepton-like mass, $m_\pi \approx m_N/7 \approx m_{\rm muon}$, seemingly impervious;
(\emph{b}) Given that the proton is (somehow) composed of valence quarks and gluon and quark partons, how is its measurable $J=1/2$ spin distributed amongst these constituents;
and (\emph{c}) Accepting that QCD's non-Abelian character is expressed as self-interactions amongst gluon partons, what impacts do these self-interactions have on both gluons themselves and measurable quantities?
It is expected that data obtained from an array of modern and anticipated facilities will open the doors to solutions \cite{Ent:2015kec, Denisov:2018unjF, Aguilar:2019teb, Andrieux:2020, Roberts:Strong2020, Anderle:2021wcy, Arrington:2021biu, Quintans:2022utc, Wang:2022xad, Accardi:2023chb, Achenbach:2025kfx, Lu:2025bnm, Messchendorp:2025men}.
The answers are important because this basic mass unit, $m_N$, emerged $\approx 1\,\mu$s after the Big Bang and thereafter had a critical influence on Universe evolution.

One may trace the Lagrangian of QCD back to Ref.\,\cite{Fritzsch:1973pi}, wherein it was stated:
``\emph{The quarks come in three `colors,' but all physical states and interactions are supposed to be singlets with respect to the SU$(3)$ of color. Thus, we do not accept theories in which quarks are real, observable particles; nor do we allow any scheme in which the color non-singlet degrees of freedom can be excited}.''
Thus, from the outset, strong-interaction practitioners were presented with a raft of basic challenges.
If gluons and quarks are unobservable, then what are the asymptotic (detectable) degrees-of-freedom (dof);
and how are they constructed from the Lagrangian dof?
Moreover, if the dof used to compose the QCD Lagrangian are not directly measurable, then essentially nonperturbative dynamics underlies all observable strong interaction physics.
In such circumstances, what theory tools can be used to test whether QCD is really the theory of strong interactions?
Ultimately and even more fundamentally, since QCD is a Poincar\'e-invariant quantum gauge field theory in four dimensions, is it truly a theory or simply one more of humankind's effective field theories?
The answer to this last question may have implications that stretch far beyond the SM.

The process of laying the path between QCD predictions and existing and expected data should be objective, \emph{viz}.\ independent of measurement process, reference frame, and other observer-dependent (subjective) positions \cite{Brodsky:2022fqy}.
Owing to the manifestly nonperturbative character of QCD, this requirement places heavy demands on phenomenology and theory, neither of which is yet fully able to satisfy them; ergo, today, there is much debate over interpretations of even extant data and their usefulness as means by which QCD can be tested.

\section{Emergent Hadron Mass (EHM) Paradigm}
An objective component decomposition of $m_N$ contains only three distinct contributions.
[\emph{I}] Higgs boson alone.
This piece is the fraction of $m_N$ that can unequivocally be attributed to the renormalisation group invariant (RGI) quark current masses generated by Higgs boson couplings into QCD:
\begin{equation}
\hat m_H = 2\hat m_u + \hat m_d \approx 13\,{\rm MeV} = 0.014\,m_N\,.
\end{equation}
This is clearly much less than the target value.
(The RGI current mass concept is as old as QCD \cite{Politzer:1976tv}.)
[\emph{II}] Constructive interference between this component and Higgs-unrelated mass generating mechanisms.
QCD-connected theory or experimental data can be used to determine the size of the in-nucleon expectation value of QCD's RGI mass term \cite{Flambaum:2005kc, Alarcon:2021dlz}:
\begin{equation}
\hat \sigma_N := \langle N(k) | \int d^4 x\, \hat m [\widehat{\bar q(x) q(x)}] |N(k)\rangle \approx 60\,{\rm MeV}\,.
\end{equation}
So, constructive interference between Higgs-alone and other sources of mass yields
\begin{equation}
\hat m_\sigma = \hat \sigma_N - \hat m_H
\approx (60-13)\,{\rm MeV} = 0.05\,m_N\,.
\end{equation}
With [\emph{I}] and [\emph{II}], one still remains far from the desired value.
[\emph{III}] EHM is the missing component, \emph{viz}.
\begin{equation}
\hat m_{\rm EHM} \approx 0.94\,m_N\,.
\end{equation}
Evidently, the challenge posed in (\emph{a}) above is either to reveal the QCD origin of
\begin{equation}
\hat m_\sigma + \hat m_{\rm EHM} \approx 0.99 m_N\,,
\end{equation}
or discard QCD as the underlying theory of strong interactions.
\emph{N.B}.\ Since the nucleon contains zero valence $s$ quarks, then any $s$-quark $\hat\sigma_N$-like term is properly identified as part of $\hat m_{\rm EHM}$.

The seed for a QCD explanation of EHM was planted in Ref.\,\cite{Cornwall:1981zr}.
Namely, owing to strong self-interactions, enabled by QCD's $3$- and $4$-gluon vertices, gluon partons metamorphose into gluon quasiparticles, whose propagation is modulated by a RGI momentum-dependent mass function that is large at infrared momenta.
The associated RGI gluon mass scale, $\hat m = 0.43(1)\,$GeV \cite{Binosi:2016nme, Cui:2019dwv, Brodsky:2024zev}, is roughly half the proton mass.
This phenomenon lies at the core of the modern EHM paradigm \cite{Roberts:2021nhw, Binosi:2022djx, Ding:2022ows, Roberts:2022rxm, Raya:2024ejx, Ferreira:2023fva, Carman:2023zke}.

The emergence of $\hat m \approx m_N/2$ has many consequences.  For instance, consider the QCD running coupling and some of the questions posed in Ref.\,\cite{Dokshitzer:1998nz}:
``\emph{We are used to the notion of a running QCD coupling.  What is its origin and status?  Formally, $\alpha_s$ is a parameter of the perturbative expansion.  From this point of view, its choice is, to a large extent, a matter of free will: it depends on how smart we want to be in organising the [perturbation theory] PT-series.  It starts to make more sense, and becomes natural, within a programme of trading a formal expansion parameter for a smarter object, a momentum-dependent $\alpha_s(k^2)$, which embodies some specific all-order radiative correction effects and is supposed to truly represent the interaction strength at a given momentum scale}.''
The existence of the gluon RGI running mass enables one to define and calculate a unique analogue in QCD of the Gell-Mann--Low effective charge in quantum electrodynamics (QED) \cite{GellMann:1954fq}.
This archetype of all running couplings is now known with a precision that locates the uncertainty in the twelfth significant figure \cite{ParticleDataGroup:2024cfk}.
The QCD effective charge \cite{Binosi:2016nme, Cui:2019dwv, Brodsky:2024zev}, $\hat\alpha(k^2)$, is process-independent (PI), just like the QED running coupling.
Furthermore, it has many additional properties, \emph{inter alia}, conformal behaviour at infrared momenta, with $\hat\alpha(k^2=0)/\pi=0.97(4)$ -- see, \emph{e.g}., Ref.\,\cite[Fig.\,3]{Ding:2022ows} and associated discussion, which make it an ideal candidate for the long-sought ``smarter object'' which represents the QCD interaction strength all length scales.

The running gluon mass and PI effective charge combine to form a principal component of the kernel in the quark gap (Dyson) equation.
The completing piece is defined by the gluon + quark vertex \cite{Binosi:2016wcx}.
The gap equation built from these elements yields a solution that expresses the third (matter) pillar of EHM, \emph{i.e}., the RGI running quark mass, $M(k^2)$; see, \emph{e.g}., Ref.\,\cite[Fig.\,2.5]{Roberts:2021nhw}.
So, the QCD Lagrangian quark partons are also transmogrified, becoming dressed-quark quasiparticles characterised, too, by a dynamically-generated RGI mass function.
For light quarks, one finds $M(0) \approx m_N/3$.
In this result, via the Faddeev equation approach to the nucleon bound-state problem \cite{Eichmann:2009qa, Yao:2024ixu}, the origin of the nucleon mass scale in QCD is immediately apparent.
However, the near masslessness of the quark + antiquark pion is not; thus in attempting to answer (\emph{a}), it is to $m_\pi$ that one must now turn.

\section{Light Pseudoscalar Mesons}
\label{Sec3}
Beginning with QCD's axialvector Ward-Green-Takahashi identity, one can prove that the mass-squared, $m_{\mathsf P}^2$, of each pion- and kaon-like pseudoscalar meson in the SU$(3)$-flavour octet can be obtained from \cite{Maris:1997tm}:
\begin{equation}
m_{\mathsf P}^2  f_{\mathsf P} = (\hat m_{\mathsf P}^1 + \hat m_{\mathsf P}^2)  \hat \kappa_{\mathsf P}\,,
\label{GMORM}
\end{equation}
where:
$\hat m_{\mathsf P}^{1,2}$ are the RGI current masses of the meson's valence quarks;
$f_{\mathsf P}$ is the meson's leptonic decay constant, \emph{viz}.\ the pseudovector projection of the bound-state's wave function onto the origin in configuration space;
and $\hat \kappa_{\mathsf P}$ is the pseudoscalar analogue of $f_{\mathsf P}$ \cite{Chang:2013epa}.
One can furthermore prove that after turning off Higgs boson couplings into QCD, $f_{\mathsf P} \neq 0 \neq \hat \kappa_{\mathsf P}$ if, and only if, chiral symmetry is dynamically broken and, consequently, these pseudoscalar mesons emerge as massless bound states of the given, defining valence quark + antiquark pair.
Microscopically, pions and kaons are (near) Nambu-Goldstone bosons because the dressing mass acquired by the valence quasiparticle constituents is (almost) precisely cancelled by the energy of interaction between them \cite{Bender:1996bb, Roberts:2016vyn, Roberts:2020udq}.
This cancellation is forbidden to $\eta$, $\eta^\prime$ mesons by the non-Abelian anomaly \cite{Bhagwat:2007ha}.

Owing to a set of Goldberger-Treiman identities that emerge in proving Eq.\,\eqref{GMORM} -- see, \emph{e.g}., Ref.\,\cite[Eqs.\,(34--37)]{Maris:1997tm} -- one finds that pion properties present the cleanest expressions of EHM mechanisms in Nature.  Some of the observable impacts are discussed in Refs.\,\cite{Roberts:2021nhw, Binosi:2022djx, Ding:2022ows, Roberts:2022rxm, Raya:2024ejx}.
It is worth illustrating this by sketching EHM expressions in the pion elastic electromagnetic form factor, $F_\pi(Q^2)$.
(Kaons and flavour-kindred heavier mesons introduce the ability to probe the [\emph{II}] components of hadron masses, \emph{i.e}.,  EHM + Higgs boson interference \cite{Ding:2022ows}.)

Crucially, science is entering a watershed period because it is becoming possible to test robust pion structure predictions, which are being delivered by modern theory \cite{Roberts:2021nhw, Binosi:2022djx, Ding:2022ows, Roberts:2022rxm, Raya:2024ejx}, using high-luminosity experimental facilities.
Such centres are in operation \cite{Ent:2015kec, Denisov:2018unjF, Andrieux:2020, Roberts:Strong2020, Quintans:2022utc}, under construction \cite{Aguilar:2019teb, Arrington:2021biu}, or being planned \cite{Anderle:2021wcy, Wang:2022xad, Accardi:2023chb}.
They (will) enable experiment to work with what are effectively $\pi$ targets by exploiting Drell-Yan reactions \cite{Denisov:2018unjF, Aguilar:2019teb, Andrieux:2020, Roberts:Strong2020, Quintans:2022utc} or the Sullivan process \cite{Ent:2015kec, Arrington:2021biu, Anderle:2021wcy, Wang:2022xad, Accardi:2023chb, Lu:2025bnm}.
Experiments on ``$\pi$ ($K$) targets'' are otherwise impossible owing to the short lifetimes of these states.

First precise pion electromagnetic form factor data (beyond the charge radius domain) were obtained (Sullivan process) at Jefferson Laboratory (JLab) around 25 years ago \cite{Volmer:2000ek}.
Today, nine precise data are available \cite{Horn:2007ug, Huber:2008id}, reaching out to $Q^2 \approx 2.5\,$GeV$^2$; and, within one year, JLab is expected to deliver additional data that push coverage to $Q^2 \approx 9\,$GeV$^2$.
Moreover, high-profile experiments are being developed for electron ion colliders \cite{Aguilar:2019teb, Arrington:2021biu, Wang:2022xad, Lu:2025bnm}.
It is reasonable to ask: Why?

The answer is a striking predictions about pseudoscalar meson structure that was delivered more than 40 years ago.  Namely \cite{Lepage:1979zb, Efremov:1979qk, Lepage:1980fj}: for a positive-charge pseudoscalar meson $\mathsf P$, constituted from valence $f$, $\bar g$ quarks,
{\allowdisplaybreaks
\begin{subequations}
\label{eq:pionFFUV}
\begin{align}
\exists \, Q_0 \gg  m_N & \; |  \;
Q^2 F_{\mathsf P}(Q^2) \stackrel{Q^2 > Q_0^2}{\approx} 16 \pi \alpha_s(Q^2)  f_{\mathsf P}^2 w_{\varphi_{\mathsf P}}^2(Q^2) \,, \\
w_{\varphi_{\mathsf P}}^2 & = e_f w_{\varphi_{{\mathsf P}^f}}^2 + e_{\bar g} w_{\varphi_{{\mathsf P}^{\bar g}}}^2 \\
w_{\varphi_{{\mathsf P}^f}} & = \frac{1}{3} \int_0^1 dx\, \frac{1}{x}\, \varphi_{\mathsf P}(x;Q^2) \,, \\
w_{\varphi_{{\mathsf P}^{\bar g}}} & = \frac{1}{3} \int_0^1 dx\, \frac{1}{1-x}\, \varphi_{\mathsf P}(x;Q^2) \,,
\end{align}
\end{subequations}
where
$\alpha_s(Q^2)$ is the QCD running coupling and $\varphi_{\mathsf P}(x;Q^2)$ is the meson's leading parton distribution amplitude (DA) evaluated at the hard scale of the interaction.
}

Equation~\eqref{eq:pionFFUV} is remarkable physics.
It predicts that beyond some momentum scale, $Q_0$, the pseudoscalar meson form factor is precisely known:
(\emph{i}) its magnitude (normalisation) is set by the meson leptonic decay constant, $f_{\mathsf P}$, which is an order parameter for dynamical chiral symmetry breaking, a corollary of EHM;
and (\emph{ii}) scaling violations are apparent, with a $Q^2$ dependence determined by that of the running coupling and DA evolution.
On $m_N^2/Q^2 \simeq 0$, all pseudoscalar mesons have the same DA \cite{Lepage:1979zb, Efremov:1979qk, Lepage:1980fj}:
\begin{equation}
\varphi_{\mathsf P}(x;Q^2) \stackrel{m_N^2/Q^2 \simeq 0}{\approx} \varphi_{\text{as}}(x) = 6x(1 - x)\,.
\label{phiasy}
\end{equation}
There are two outstanding issues, nonetheless:
($\mathpzc Q1$) what is the value of $Q_0$ beyond which scaling violations become apparent;
and
($\mathpzc Q2$) what is the pointwise form of $\varphi_{P}(x;Q^2)$ that is appropriate for experiments at terrestrially accessible scales?

Only a nonperturbative approach to QCD can answer questions like those raised above and continuum Schwinger function methods (CSMs) provide one such widely used tool; see, \emph{e.g}., Refs.\,\cite{Roberts:2021nhw, Binosi:2022djx, Ding:2022ows, Roberts:2022rxm, Raya:2024ejx, Achenbach:2025kfx}.
Indeed, the pion and kaon form factor problems were most recently addressed in Ref.\,\cite{Yao:2024drm}, which employed a symmetry-preserving approximation scheme for all required quantum field equations, augmented by methods from applied mathematics, to deliver predictions for $F_{\pi,K}(Q^2)$ that cover the domain $0\leq Q^2/{\rm GeV}^2\leq 30$, \emph{i.e}., the entire range of momentum transfers that may become accessible to precision experiments over the next twenty years or more \cite{Arrington:2021biu, Anderle:2021wcy, Wang:2022xad, Accardi:2023chb, Lu:2025bnm}.

\begin{figure}[t]
\vspace*{1.5em}

\includegraphics[width=0.41\textwidth]{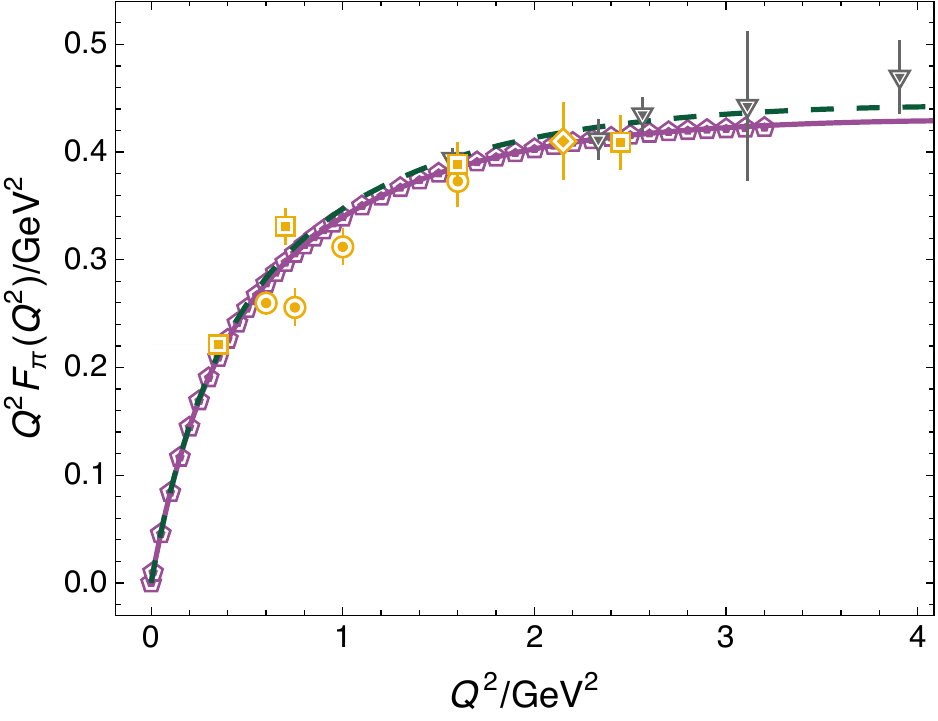}
\vspace*{-40ex}

\leftline{\hspace*{0.1em}{\large{\textsf{A}}}}

\vspace*{40ex}

\includegraphics[width=0.41\textwidth]{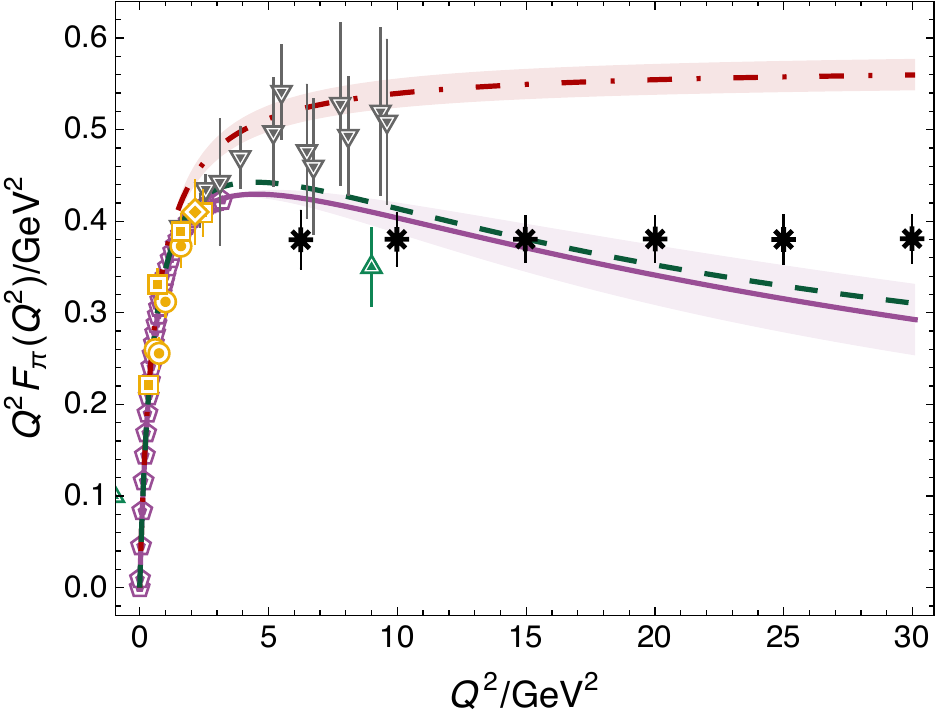}
\vspace*{-40ex}

\leftline{\hspace*{0.1em}{\large{\textsf{B}}}}

\vspace*{36ex}

\caption{\label{FigFpi}
Pion elastic electromagnetic form factor, $Q^2 F_\pi(Q^2)$.
%
Legend.
Purple curves -- CSM predictions from Ref.\,\cite{Yao:2024drm};
dashed green curve  -- CSM results obtained using perturbation theory integral representations \cite{Chang:2013nia, Gao:2017mmp};
grey down-triangles -- lattice-regularised QCD \cite{Ding:2023fac}.
Data (gold) -- diamond \cite{Horn:2007ug}; circles and squares \cite{Huber:2008id}.
Panel B only.
Dot-dashed red curve -- monopole with mass fixed by empirical pion charge radius \cite{Dally:1982zk, Amendolia:1986wj, Cui:2021aee}: $r_\pi \approx 0.66\,$fm.
Further: green up triangle -- estimated uncertainty of forthcoming JLab measurement at the highest accessible $Q^2$ point \cite{E12-19-006};
and black asterisks -- anticipated uncertainty of EIC data, whose coverage should extend to $Q^2 \approx 35\,$GeV$^2$ \cite{Aguilar:2019teb, Arrington:2021biu}.
The central magnitude of these points was chosen arbitrarily.
}
\end{figure}

The pion form factor results are displayed in Fig.\,\ref{FigFpi}.
Evidently, the low-$Q^2$ results -- Fig.\,\ref{FigFpi}\,A -- agree with extant data inferred from experiments, with the predicted charge radius, $r_\pi=0.67\,$fm,  being a good match with that determined from $Q^2\simeq 0$ data \cite{Dally:1982zk, Amendolia:1986wj, Cui:2021aee}.
Furthermore, on the domain of available JLab data \cite{Horn:2007ug, Huber:2008id}, the predicted $Q^2$ dependence agrees well.

Looking at a larger $Q^2$ domain, Fig.\,\ref{FigFpi}\,B also displays a much used phenomenological curve, \emph{viz}.\ the vector meson dominance (VMD) result obtained with $Q^2$-monopole mass chosen to reproduce the pion charge radius.
Evidently, this quark-counting-rules scaling result begins to deviate significantly from data and CSM prediction at the upper bound of available JLab measurements.
The other (grey triangle) points depicted in Fig.\,\ref{FigFpi} are those computed in a recent lattice-QCD study \cite{Ding:2023fac}: only with much improved precision could lattice-QCD begin to distinguish between the VMD model and CSM prediction.

Notably, the CSM analysis predicts the existence of a maximum in $Q^2 F_\pi(Q^2)$ on $Q^2\simeq 4.6(5)\,$GeV$^2$.  Consistent with the perturbative QCD analysis \cite{Lepage:1979zb, Efremov:1979qk, Lepage:1980fj}, scaling violation is apparent at larger $Q^2$ values:
\begin{equation}
Q^2 F_\pi(Q^2)
\stackrel{Q^2 > 5 m_N^2}{\propto}
[ 1/ \ln Q^2 ] ^{\gamma_F}\,,\; \gamma_F \approx 1.1\,.
\end{equation}
Notably, any VMD-linked model must exhibit an opposite trend, \emph{viz}.\ a result for $Q^2 F_\pi(Q^2) $ which rises steadily with increasing $Q^2$, without inflection, toward a finite ultraviolet limit.
The future may see the $Q^2$ reach and precision of lattice-QCD results come to match those of extant CSM studies and so also see the breakaway from VMD and onset of scaling violation.

Much the same things can be said about predictions for $F_K(Q^2)$ \cite{Yao:2024drm}.
Notably, unresolved controversies surround the timelike behaviour of the charged and neutral kaon elastic form factors; see, \emph{e.g}., Ref.\,\cite[Fig.\,4]{Gao:2017mmp} and associated discussion.

\section{Nucleons}
Using CSMs, the nucleon bound-state problem is typically tackled via the Poincar\'e-covariant Faddeev equation; see, \emph{e.g}., Ref.\,\cite[Fig.\,1]{Yao:2024ixu}.
The studies in Refs.\,\cite{Yao:2024ixu, Yao:2024uej} used precisely the same approach as that described in Sec.\,\ref{Sec3} to deliver a unifying set of parameter-free predictions for nucleon elastic electromagnetic and gravitational form factors and their species separations to large $Q^2$ values.
All these predictions link observables directly with fundamental QCD Schwinger functions and thereby expose novel expressions and corollaries of EHM.

\begin{figure}[t]
\vspace*{1.2em}

\leftline{\hspace*{0.5em}{\large{\textsf{A}}}}
\vspace*{-5ex}
\includegraphics[clip, width=0.9\linewidth]{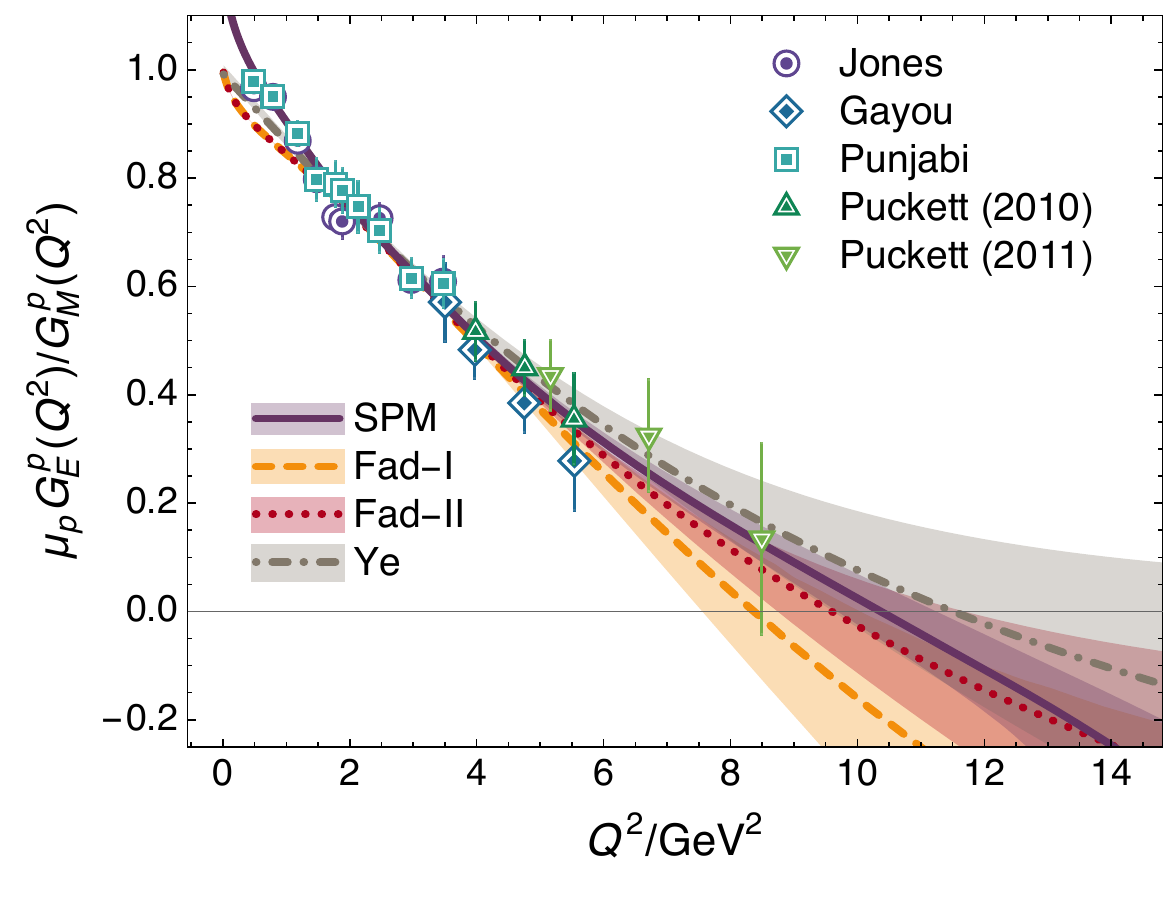}
\vspace*{1ex}
\leftline{\hspace*{0.5em}{\large{\textsf{B}}}}
\vspace*{-5ex}
\includegraphics[clip, width=0.9\linewidth]{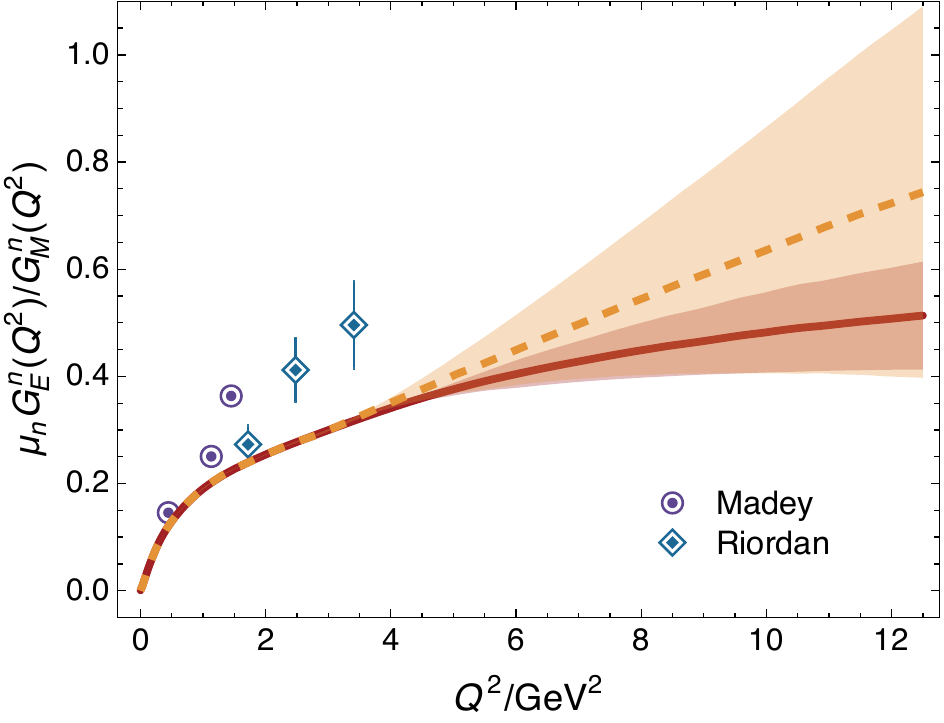}
\vspace*{4.5ex}

%
\caption{\label{FigF2}
{\sf Panel A}.  SPM prediction for $\mu_p G_E^p(Q^2)/ G_M^p(Q^2)$.
The image also depicts modern parameter-free Faddeev equation predictions \cite[Fad-I, Fad-II]{Yao:2024uej} and the result obtained via a subjective phenomenological fit to the world's electron + nucleon scattering data \cite[Ye]{Ye:2017gyb}.
%
{\sf Panel B}.  CSM predictions for $\mu_n G_E^n/G_M^n$:
Fad-I -- dashed orange curve within like-coloured band; and Fad-II -- solid red curve within like-coloured band.  (See Ref.\,\cite{Yao:2024uej} for details.)
Data: proton -- Refs.\,\cite{Jones:1999rz, Gayou:2001qd, Punjabi:2005wq, Puckett:2010ac, Puckett:2017flj}; and neutron -- Refs.\,\cite{Madey:2003av, Riordan:2010id}.
%
}
\end{figure}

The CSM progress brings within reach the study of a particular highlight of electron + proton scattering experiments this century, \emph{i.e}., data which suggests the existence of a zero in $G_E^p$, the proton elastic electric form factor \cite{Jones:1999rz, Gayou:2001qd, Punjabi:2005wq, Puckett:2010ac, Puckett:2017flj}.
(A zero in the proton $\to$ Roper transverse helicity transition amplitude has unambiguously been located \cite{Burkert:2019bhp}.  Empirically, $G_M(Q^2) > 0$ out to at least $Q^2=20\,$GeV$^2$ \cite{Ye:2017gyb}.)
Against this background and without reference to any model/theory of strong interactions, Ref.\,\cite{Cheng:2024cxk} subjected the $29$ available data on $\mu_p G_E^p(Q^2)/ G_M^p(Q^2)$ to analysis using the Schlessinger point method (SPM) \cite{Schlessinger:1966zz, PhysRev.167.1411, Tripolt:2016cya}.
This is an objective approach for extracting model-independent information, with quantified uncertainties, from a body of precise data.  The SPM is founded in analytic function theory and applicable in the same form to diverse systems and observables.
In Ref.\,\cite{Cheng:2024cxk}, $50\,000$ function-form unbiased interpolator estimates of the curve that underlies the data were developed and the following conclusions reached:
with 50\% confidence, extant $\mu_p G_E^p(Q^2)/ G_M^p(Q^2)$ data are consistent with the existence of a zero in the ratio on $Q^2 \leq 10.37\,$GeV$^2$;
the level of confidence rises to 99.9\% on $Q^2 \leq 13.06\,$GeV$^2$;
and the likelihood that the data are consistent with the absence of a zero in the ratio on $Q^2 \leq 14.49\,$GeV$^2$ is $1/1$-million.
These conclusions are expressed by the purple curve and associated band in Fig.\,\ref{FigF2}\,A.

\begin{figure*}[t]
\hspace*{-1ex}\begin{tabular}{lll}
\large{\textsf{A}} & \large{\textsf{B}} & \large{\textsf{C}}\\[-0.5ex]
%
\includegraphics[clip, width=0.33\textwidth]{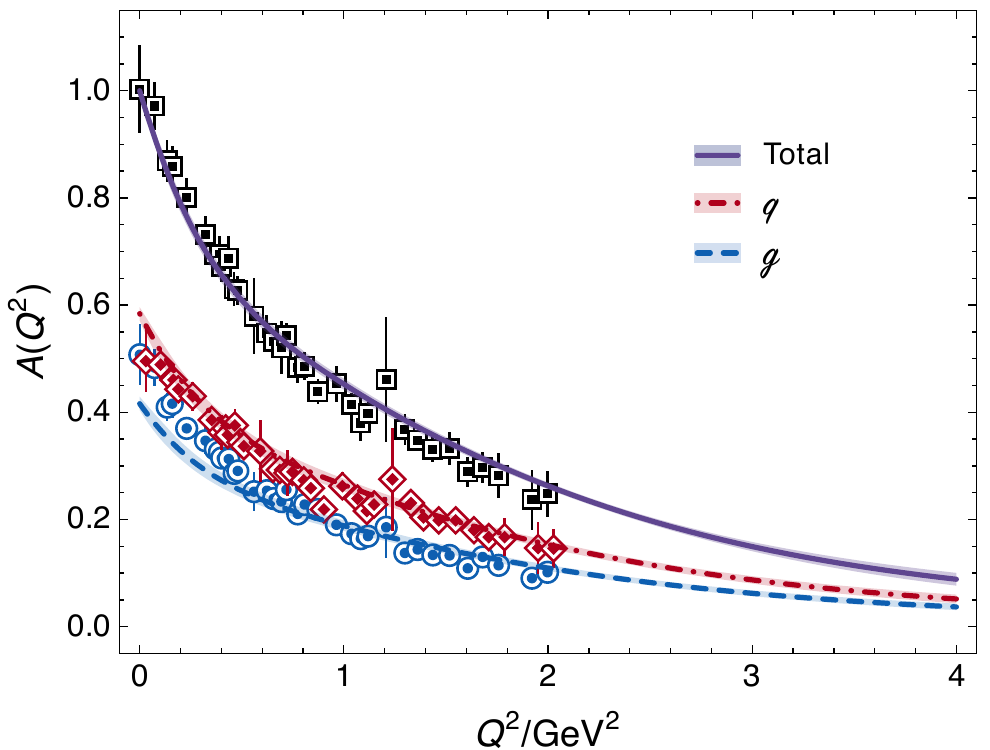} \hspace*{-1.2em} &
\includegraphics[clip, width=0.33\textwidth]{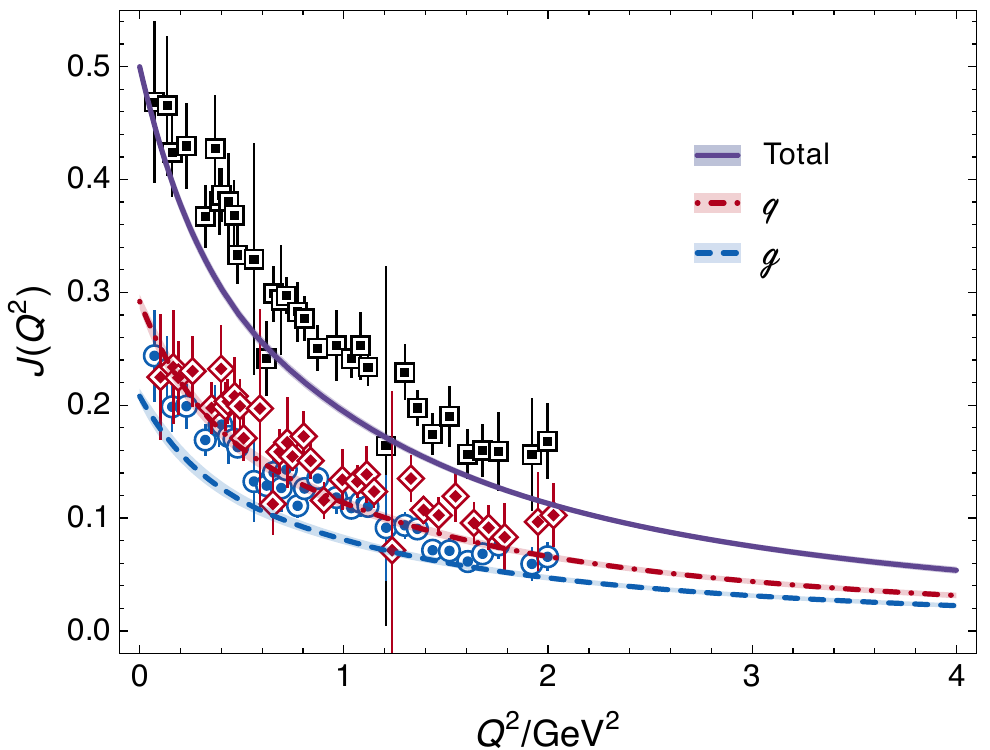} \hspace*{-1.2em} &
\includegraphics[clip, width=0.33\textwidth]{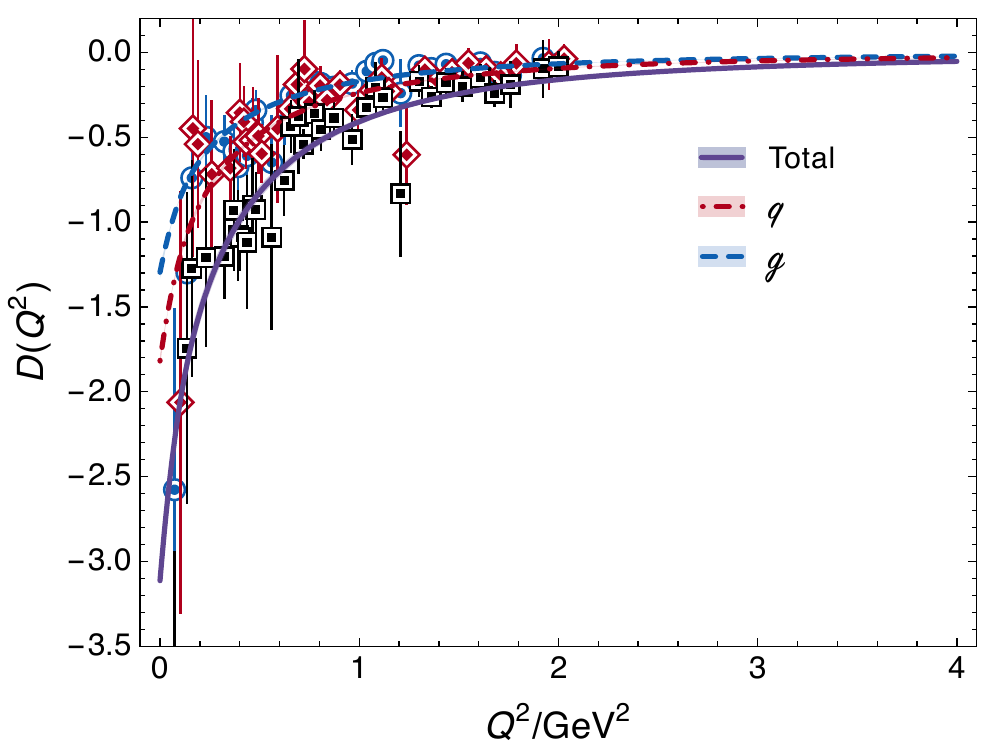}
\end{tabular}
\caption{\label{PlotGFFsAJ}
Nucleon gravitational form factors.
Curves -- CSM predictions \cite{Yao:2024ixu}.  The bracketing bands mark the extent of $1\sigma$ SPM uncertainty.
Each overall (species-summed) result is independent of resolving scale, $\zeta$; but, naturally, species decompositions evolve with $\zeta$.
Points in each panel -- lattice-QCD results reproduced from Ref.\,\cite{Hackett:2023rif}:
black squares -- total form factor;
red diamonds -- quark;
blue circles -- glue component.
}
\end{figure*}

Earlier, therefore independent of the Ref.\,\cite{Cheng:2024cxk} study, Ref.\,\cite{Yao:2024uej} employed a 
symmetry-preserving 
analysis of the quantum field equations required for calculation of hadron masses and interactions to deliver parameter-free predictions for all nucleon
charge and magnetisation distributions.
The outcomes of this study are as follows.
The proton electric form factor, $G_E^p(Q^2)$, is predicted to possess a zero at a $Q^2$ location within reach of modern experiments:
\begin{equation}
\label{ProtonZero}
Q^2 /{\rm GeV}^2= 8.86^{+1.93}_{-0.86}.
\end{equation}
The value falls easily within the domain determined empirically in Ref.\,\cite{Cheng:2024cxk}.
These things are evident in Fig.\,\ref{FigF2}, wherein the Fad-I and Fad-II curves and bands depict results obtained in Ref.\,\cite{Yao:2024uej} using two distinct methods for calculating $\mu_p G_E^p(Q^2)/ G_M^p(Q^2)$ at large $Q^2$.  The number in Eq.\,\eqref{ProtonZero} is the average of these results.

On the other hand, the neutron electric form factor, $G_E^n$, does not exhibit a zero; see Fig.\,\ref{FigF2}\,B.
Consequently, experiments, underway or planned, will see $|G_E^n/G_E^p|>1$, \emph{viz}.\  an electric form factor of the charge-neutral neutron which is greater than that of the charge-one proton.
This is predicted to occur at $Q^2/{\rm GeV}^2 = 4.66^{+0.18}_{-0.13}$.

It was further argued in Ref.\,\cite{Yao:2024uej} that these outcomes owe to the behaviour of the proton's $d$-quark Pauli form factor.

All the form factor features predicted in Ref.\,\cite{Yao:2024uej} are sensitive expressions of emergent phenomena in QCD.
Significantly, no real improvement over the Ref.\,\cite{Yao:2024uej} analysis can be anticipated before lattice-regularised QCD produces precise results on a similar domain to that discussed therein.

With the CSM framework in Ref.\,\cite{Yao:2024uej} validated by comparisons with data, it may be used to deliver solidly benchmarked predictions for the entire array of nucleon elastic form factors.  For instance, nucleon gravitational form factors (GFFs) are calculable once the dressed graviton + quark vertex is known.  That vertex was a key outcome of the study of $\pi$, $K$ electromagnetic and gravitational form factors in Ref.\,\cite{Xu:2023izo}; and it was used in Ref.\,\cite{Yao:2024ixu} to deliver parameter-free predictions for all three nonzero nucleon GFFs that characterise the expectation value of the QCD energy-momentum tensor in the nucleon:
\begin{align}
m_N \Lambda_{\mu\nu}^{Ng}(Q) & = - \Lambda_+(p_f)
[
K_\mu K_\nu A(Q^2) + i K_{\left\{\mu\right.}\!\sigma_{\left.\nu\right\}}\,\!_\rho Q_\rho J(Q^2)
 \nonumber \\
& \qquad  + \tfrac{1}{4} (Q_\mu Q_\nu - \delta_{\mu\nu} Q^2) D(Q^2)
]
\Lambda_+(p_i)  \,,
\label{EMTproton}
\end{align}
where
$p_{i,f}$ are the momenta of the incoming/outgoing nucleon, $p_{i,f}^2=-m_N^2$,
$K=(p_i+p_f)/2$, $Q=p_f-p_i$;
all Dirac matrices are standard \cite[Sec.\,2]{Roberts:2000aa}, with $\sigma_{\mu\nu}= (i/2)[\gamma_\mu,\gamma_\nu]$;
$\Lambda_+$ is the projection operator associated with a positive energy nucleon;
and $a_{\left\{\mu\right.}\!b_{\left.\nu\right\}}=(a_\mu b_\nu + a_\nu b_\mu)/2$.

Each form factor in Eq.\,\eqref{EMTproton} is Poincar\'e invariant; hence, observable:
$A$ is the nucleon mass distribution form factor;
$J$ describes the distribution of total angular momentum;
and
$D$ expresses in-nucleon pressure and shear forces.
In the forward limit, $p_f=p_i$, symmetries entail $A(0)=1$, $J(0)=1/2$.
$D(0)$ is also a conserved charge; however, akin with the nucleon axial charge, $g_A$, its value is a dynamical property of the nucleon
Having been described as ``\ldots \emph{the last unknown global property of the nucleon} \ldots'' \cite{Polyakov:2018zvc}, it is a focus of much attention.

CSM predictions for the proton GFFs \cite{Yao:2024ixu} are displayed in Fig.\,\ref{PlotGFFsAJ}.
The symmetry preserving character of the analysis is evident in the values $A(0)=1$, $J(0)=1/2$.
Furthermore, in delivering $J(0)=1/2$, the study confirms that the anomalous gravitomagnetic moment of a spin-$1/2$ system is zero \cite{Kobzarev:1962wt, Teryaev:1999su, Brodsky:2000ii}.
The study predicted \cite{Yao:2024ixu}:
\begin{equation}
D(0) = -3.11(1)\,.
\end{equation}
This value was subsequently confirmed by a data-informed extraction \cite{Cao:2024zlf}: $D(0) = -3.38_{-0.32}^{+0.26}$.
The value of the analogous quantity in the pion is $[-\theta_1(Q^2=0)] = -0.97$ \cite{Xu:2023izo}.

Working with the form factors in Eq.\,\eqref{EMTproton}, one constructs the (observable) proton mass-energy density form factor as follows:
\begin{equation}
{\cal M}(Q^2) =
A(Q^2) + \frac{Q^2}{4 m_p^2} [A(Q^2)-2 J(Q^2)+D(Q^2)]\,.
\label{MQ2}
\end{equation}
The CSM study predicts that ${\cal M}(Q^2)$ is positive definite on a domain that extends, at least, to $Q^2=100\,$GeV$^2$.

${\cal M}(Q^2) $ is a direct analogue of the nucleon electric charge distribution that is measured in electron scattering.  Naturally, since it is  Poincar\'e-invariant, then it does not matter whether one builds a spatial density using a two- or three-dimensional Fourier transform.  No projective mapping can yield new information from a Poincar\'e-invariant source function and any interpretation of the mapping's outcome will always be subjective (practitioner dependent).
As usual, the associated mass-energy radius may be obtained via
%
\begin{equation}
\label{MRadius}
\langle r^2\rangle_{\rm mass}  = -6 \left. \frac{d}{dQ^2} A(Q^2) \right|_{Q^2=0}
- \frac{3}{2m_p^2}D(0)\,.
\end{equation}
This expression has the same form as that for the proton electric charge radius; so, it is a comparison between these two quantities that is natural.  In both cases, the radius increases as the magnitude of the symmetry-unconstrained pressure/anomalous-magnetic-moment term increases and with the same rate.
A ``mechanical'' radius can also be obtained via the normal force distribution form factor:
\begin{equation}
\langle r^2\rangle_{\rm mech} = 6 \bigg/ \int_0^\infty dt\,[D(t=Q^2)/D(0)]\,.
\end{equation}
The CSM analysis predicts \cite{Yao:2024ixu}:
\begin{equation}
\label{RadOrder}
r_{\rm mass} = 0.81(5) r_{\rm ch} > r_{\rm mech} = 0.72(2) r_{\rm ch}\,,
\end{equation}
where $r_{\rm ch}=0.887(3)\,$fm is the proton charge radius calculated using the same framework \cite{Yao:2024uej}.

\emph{N.B}.\ In Ref.\,\cite{Yao:2024uej}, no attempt was made to reproduce the currently accepted value of $r_{\rm ch}=0.84\,$fm \cite{Cui:2021skn, Cui:2022fyr, Xiong:2023zih}.  A $\sim 5$\% overestimate was judged acceptable in that first \emph{ab initio} unification  of nucleon and meson form factors.  The overestimate has no qualitative impact on the ordering in Eq.\,\eqref{RadOrder}.

It is worth stressing that in opposition to the complete, Poincar\'e-invariant, measurable form factors discussed above, any decomposition of such form factors into distinct species contributions is frame- and scale-dependent; hence, subjective (practitioner dependent).
At the hadron scale \cite{Yin:2023dbw}, denoted $\zeta_{\cal H}$, the properties of each form factor are completely determined by those of the dressed valence quark dof and the binding between them: all partonic glue and sea-quark contributions are sublimated into the dressed valence quarks and interaction.
This is the essence of ``dressing'', which is expressed in any QCD Schwinger function -- consider, \textit{e.g}., a diagrammatic (or series) expansion of the quark gap equation \cite{Mattuck:1976xt, Roberts:1994dr}.
It follows that no comparison between a proton charge or magnetic radius and an in-proton quark-flavour radius or gluon radius is objectively valid: one simply cannot meaningfully compare a Poincar\'e-invariant observable with a frame- and scale-dependent quantity.
Even a comparison between quark and glue radii is subjective because the values change with probe resolving scale.

\begin{figure}[t]


\centerline{%
\includegraphics[clip, width=0.9\linewidth]{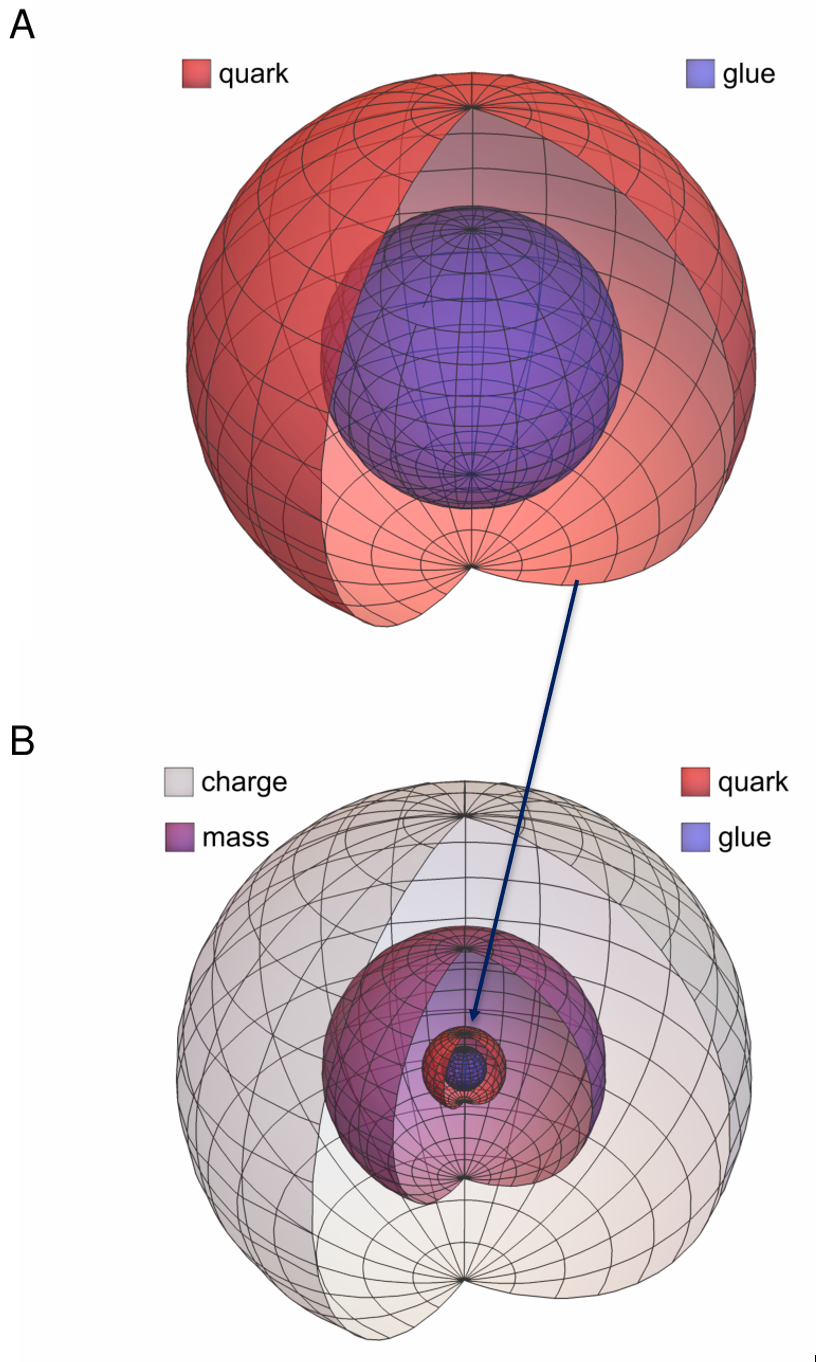}}

\caption{\label{ratioplotW}
{\sf Panel A}.
Sketch representing the relative sizes of the in-proton spacetime volumes occupied by quark and gluon mass-energy distributions at resolving scale $\zeta_2$.
{\sf Panel B}.  Combined as required, the scale-dependent distributions in the upper panel yield the scale-invariant (observable) mass-energy distribution (purple 3/4-sphere) in this image.  Plainly, the spacetime volume occupied by the proton mass-energy distribution is far smaller than that occupied by the charge distribution (grey 3/4-sphere).
(Images based on information in Eqs.\,\eqref{RadOrder}, \eqref{z2ratio}.)
}
\end{figure}

One may nevertheless discuss species decompositions so long as the context is clear and any comparisons are between kindred quantities.  Hence, using the all-orders evolution scheme \cite{Yin:2023dbw}, Ref.\,\cite{Yao:2024ixu} delivered a species decomposition.
It was found that for each form factor, the scale-dependent gluon:total-quark contribution ratio is a fixed number, \emph{viz}.\ a constant, independent of $Q^2$.  At $\zeta=\zeta_2:=2\,$GeV:
\begin{equation}
{\mathpzc g}(Q^2)/{\mathpzc q}(Q^2) = 0.71(4)\,.
\label{gonqconstant}
\end{equation}

It is critical to test the prediction in Eq.\,\eqref{gonqconstant} because it entails that no species decomposition contains any new information:
the scale-invariant observable form factors already express everything that can be known.
For instance, the relative contributions of gluons and quarks to the proton mass radius are just given by
$\langle r^2\rangle_{\mathpzc p = \mathpzc g, \mathpzc q}^\zeta = \langle x \rangle_{\mathpzc p}^\zeta r_{\rm mass}^2$, \emph{i.e}., the products of their scale-dependent light-front momentum fractions with the observable radius.
Using this relationship, one finds:
\begin{subequations}
\label{z2ratio}
\begin{align}
r_{\rm mass}^{\mathpzc g} & = 0.84(10) r_{\rm mass}^{\mathpzc q}\,,\\
r_{\rm mass}^{\mathpzc q} & =0.62(4) r_{\rm ch}\,, \label{chargeradius}
\end{align}
\end{subequations}
namely, at $\zeta_2$, gluons occupy only roughly one-half of the spacetime volume occupied by quarks.  This outcome is visualised in Fig.\,\ref{ratioplotW}\,A.
For comparison, the relative spacetime size of the charge distribution is displayed in Fig.\,\ref{ratioplotW}\,B.

\begin{figure}[t]

\includegraphics[clip, width=0.41\textwidth]{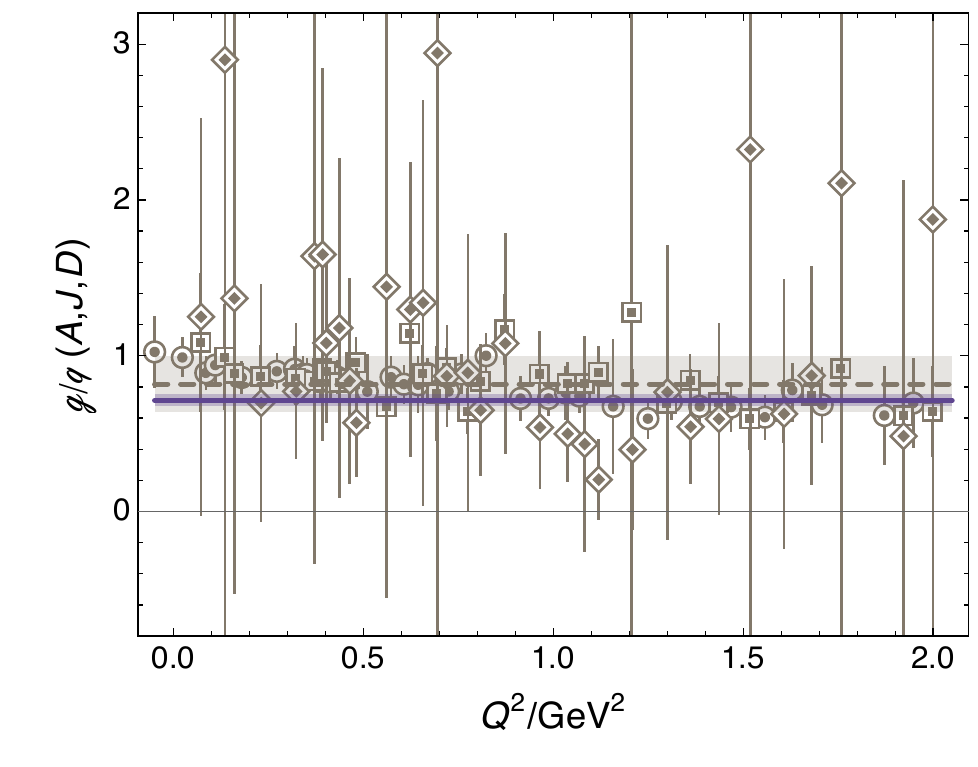}

\caption{\label{RatioCheck}
Combined $A$, $J$, $D$ lattice-QCD ${\mathpzc g}/{\mathpzc q}$ results (grey points): grey line -- uncertainty weighted average of all lattice points; and grey band -- $1\sigma$ around the central value: $0.82(18)$.
CSM prediction in Eq.\,\eqref{gonqconstant} -- solid purple line and band.
Resolving scale: $\zeta = \zeta_2$.
Separated function comparisons are drawn in Ref.\,\cite[Fig.\,5]{Yao:2024ixu}.
}
\end{figure}

Currently, the only Eq.\,\eqref{gonqconstant} comparisons available are those enabled by the lattice-QCD results in Ref.\,\cite{Hackett:2023rif}: they have large statistical uncertainties and their systematic errors have neither been quantified nor controlled.  Notwithstanding, within this uncertainty context, the lattice-QCD results are consistent with the Eq.\,\eqref{gonqconstant} prediction: ${\mathpzc g}/{\mathpzc q}_{\,\mbox{\scriptsize lattice-QCD}} = 0.82(18)$.  This is illustrated in Fig.\,\ref{RatioCheck}.

In-nucleon pressure and shear-force density profiles may be computed from $D(Q^2)$.
This is why that function is called the pressure (\emph{Druck}) form factor.
(The pseudoscalar meson analogue is often denoted $\theta_1(Q^2)$ \cite{Xu:2023izo}.)
The CSM predictions for such profiles are drawn in Ref.\,\cite[Fig.\,3]{Yao:2024ixu}.
Reviewing those predictions, one sees that the pion near-core pressure is approximately twice that in the proton.
This is natural because the pressure may loosely be connected with the modulus-squared of the bound-state's wave function.
CSM studies predict that QCD dynamics are the same within every hadron; hence,
\begin{equation}
P_\pi \approx \frac{[r_{\rm ch}^{\rm proton}]^2}{[r_{\rm ch}^{\pi}]^2} P_{\rm proton}
\approx 2 P_{\rm proton}\,.
\end{equation}
Given that the in-proton near-core pressure is ten-times greater than that in a neutron star \cite{Ozel:2016oaf}, then these hadrons, which practically fill our bodies, are the densest known systems in the Universe, excepting black holes.


It is worth highlighting that interpretations of the pressure distributions obtained from $D(Q^2)$ are qualitatively no different from those connected with the distributions inferred from the mass and spin form factors.
At $\zeta_{\cal H}$, the pressure expresses forces of attraction/repulsion and shear stress within the bound-state formed by the quasiparticle constituents.
Hence, there is a clean analogy with realisable two- and three-body systems in quantum mechanics.
At higher resolving scales, $\zeta$, reached via evolution, one may view a hadron interior as a dense partonic medium; then, the pressure expresses pairwise forces between in-medium test elements.
Each such element contains a number of partons fixed by its volume $\times$ the associated species light-front wave function magnitude-squared at its location.
Plainly, therefore, the pressure distribution form factor and its interpretations are qualitatively equivalent to the distributions of mass and spin, and of charge, magnetisation, etc.

Some practitioners imagine that objective access to the in-proton expectation value of the trace of QCD energy momentum tensor, $\langle p(k_2) | T_{\mu\nu}(Q)| p(k_1)\rangle$, is possible via photoproduction of (heavy) vector mesons from the proton because such mesons and the proton have no valence dof in common.
With this motivation, extensive new data have become available on $\gamma+p \to J/\psi + p$ and numerous related model calculations completed, in many cases with an erroneous view -- see above -- to extracting a gluon parton density and comparing it with the proton charge radius.
Detailed discussions are provided, \emph{e.g}., in Refs.\,\cite{Du:2020bqj, Lee:2022ymp, JointPhysicsAnalysisCenter:2023qgg, Tang:2024pky, Sakinah:2024cza, Tang:2025qqe}.

In association with the in-proton expectation value of the QCD trace anomaly, one may consider the following combination of observable form factors:
\begin{equation}
\theta(Q^2) =
A(Q^2) + \frac{Q^2}{4 m_p^2} [A(Q^2)-2 J(Q^2) + 3 D(Q^2)]\,.
\label{theta}
\end{equation}
Plainly, the only difference from Eq.\,\eqref{MQ2} is $D \to 3 D$; so, this compound form factor cannot contain objective information beyond that already expressed by the mass-energy form factor.

Again, $\theta(Q^2)$ is an observable.
So, like CSM results for hadron electromagnetic form factors, benchmarked against experiment, see, \emph{e.g}., Fig.\,\ref{FigF2}, and all the component form factors of $\theta(Q^2)$, drawn in Fig.\,\ref{PlotGFFsAJ}, the behaviour of this ``trace anomaly'' form factor at the hadron scale is completely determined by that of the dressed valence quark dof.
Hence, any suggestion that $\theta(Q^2)$ -- an observable -- expresses something about gluon contributions to the nucleon mass is erroneous: the very definition of ``glue'' is scale dependent.
Stated differently \cite{Roberts:2016vyn}: whilst in the deep ultraviolet one may use glue parton field variables to express $T_{\mu\mu}(Q)$, operator mixing with $m \bar q q$ means that strength shifts between these terms as the scale is reduced to infrared values, whereat the latter absorbs the former and one has a nucleon built from complex dressed-quark quasiparticles, with infrared-large momentum-dependent masses, into which the parton dof have been absorbed.


\begin{figure}[t]
\leftline{\hspace*{0.5em}{\large{\textsf{A}}}}
\vspace*{-1ex}

\includegraphics[clip, width=0.90\linewidth]{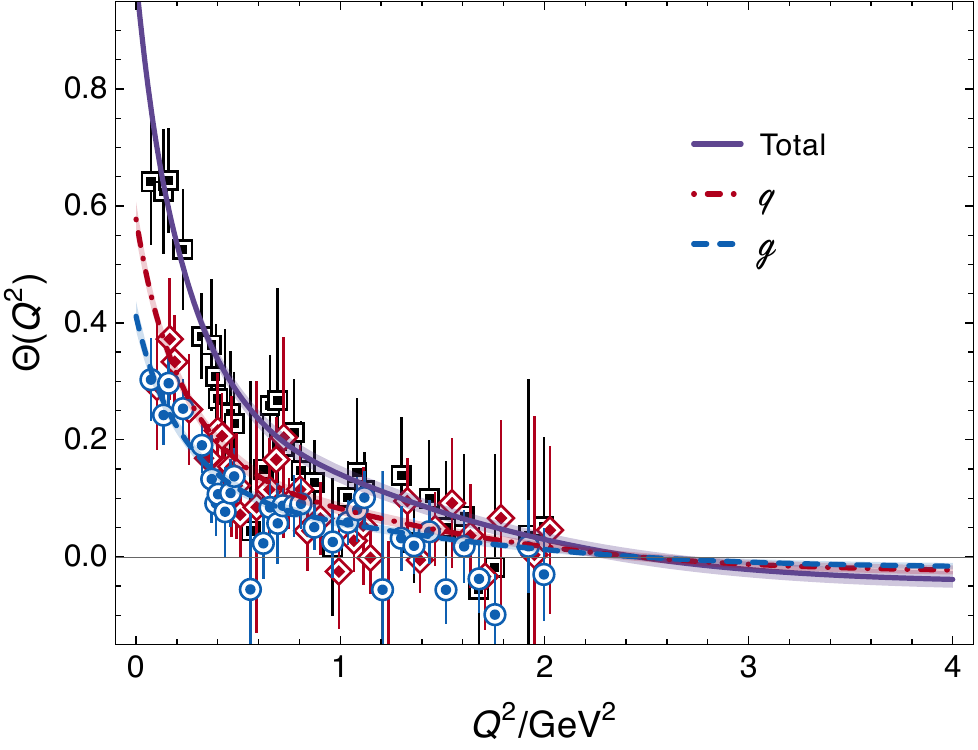}

\leftline{\hspace*{0.5em}{\large{\textsf{B}}}}
\vspace*{-1ex}
\includegraphics[clip, width=0.90\linewidth]{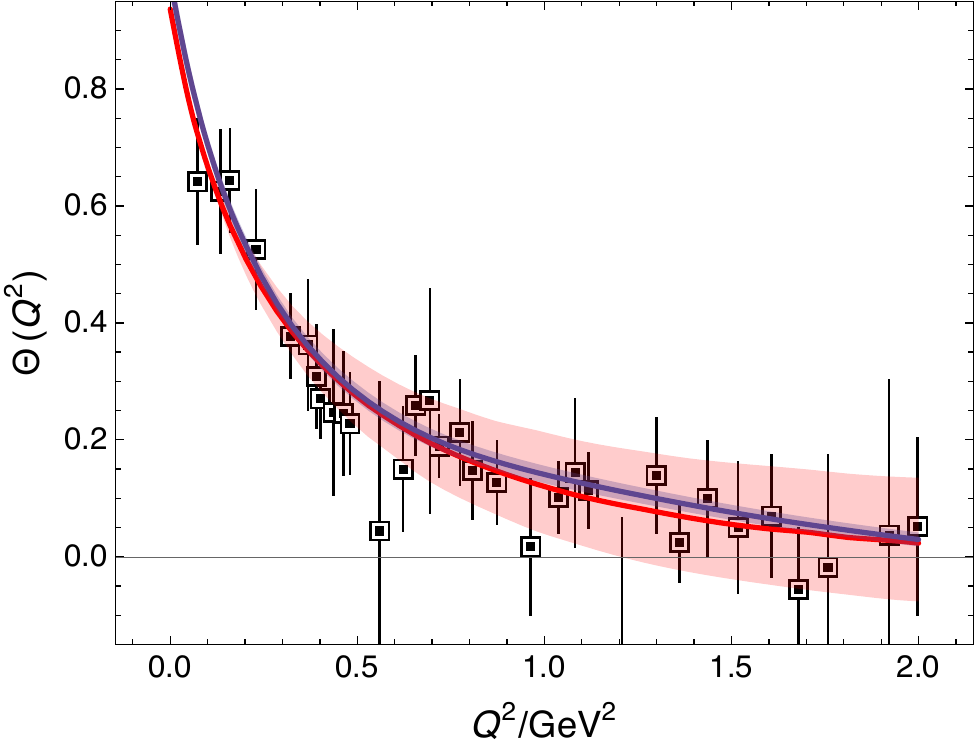}

\caption{\label{thetaQ2}
{\sf Panel A}.
    Curves -- CSM predictions, with bracketing bands that mark the extent of $1\sigma$ uncertainty in the numerical procedure for reaching large $Q^2$.
    In each case, the overall (species-summed) result is independent of resolving scale, $\zeta$.
    The species decompositions evolve with $\zeta$.
    Lattice-QCD points in each panel are obtained employing the results in Ref.\,\cite{Hackett:2023rif}:
    black squares -- total form factor;
    red diamonds -- quark component;
    blue circles -- glue.
{\sf Panel B}.
    $\theta(Q^2)$.
    CSM prediction -- purple curve and band.
    Data-based inference \cite{Cao:2024zlf} -- red curve and band.
}
\end{figure}

Figure~\ref{thetaQ2}\,A depicts the CSM prediction for $\theta(Q^2)$ \cite{Binosi:2025kpz} along with the species separation delivered by AO evolution \cite{Yao:2024ixu}.
Evidently, within the large lattice uncertainties, there is agreement between lattice-QCD and the CSM results in every case.  This agreement validates the CSM perspective sketched above.

Employing the CSM results in Fig.\,\ref{thetaQ2}\,A, one finds
\begin{equation}
    \langle r^2 \rangle_{\theta}=
    (0.960(36)\,{\rm fm})^2 = (1.08(4) r_{\rm ch})^2\,,
    \label{rtheta2}
\end{equation}
\emph{i.e}., a prediction which is confirmed by the data-informed extraction in Ref.\,\cite{Cao:2024zlf}: $(0.97^{+0.02}_{-0.03}\,{\rm fm})^2$.

Regarding Eq.\,\eqref{rtheta2}, the objective interpretation is that one chosen combination of proton gravitational form factors has a larger slope than another particular combination of electromagnetic form factors.
As stressed above, this outcome cannot be used to argue that gluons are distributed over a larger spacetime volume than quarks.
Adding to this, Fig.\,\ref{thetaQ2}\,A also presents a species decomposition of $\theta(Q^2)$.
As with ${\cal M}(Q^2)$, at a scale of $\zeta = 2\,$GeV, the quark contribution to $\langle r^2 \rangle_{\theta}$ is $\approx 40$\% larger than the gluon contribution: $\langle r^2 \rangle_{\theta}^{\mathpzc q} \approx 1.4 \langle r^2 \rangle_{\theta}^{\mathpzc g}$.

Figure~\ref{thetaQ2}\,B displays a comparison between the CSM prediction for $\theta(Q^2)$ \cite{Binosi:2025kpz} and the result inferred from data in Ref.\,\cite{Cao:2024zlf}.
Evidently, the agreement is good: on the displayed domain, the two distinct curves match almost within line-width.

A troubling feature, revealed by Fig.\,\ref{thetaQ2}\,B, is that $\theta(Q^2)$ is not positive definite.
The species decompositions exhibit the same property, see Fig.\,\ref{thetaQ2}\,A.
This is an unwelcome quality to be connected with a global property of a nucleon mass distribution and equally disturbing for the species decompositions.
Whilst charge distributions involve positive and negative charge carriers, hence can have domains of negative support; all physical mass is nonnegative, so there is seemingly no sound reason why a mass form factor should become negative.
The mass-energy form factor in Eq.\,\eqref{MQ2} is positive definite.
The problem arises because of the change $1 D(Q^2) \to 3 D(Q^2)$ in shifting from Eq.\,\eqref{MQ2} to Eq.\,\eqref{theta}.
Equation~\eqref{MQ2} is to be preferred because ${\mathpzc M}(Q^2)$ is more readily comparable with nucleon electric form factors; see, \emph{e.g}., Eq.\,\eqref{MRadius}.

It is worth recording that Ref.\,\cite{Yao:2024ixu} also included a species decomposition of the nucleon pressure form factor.
Focusing on light quarks alone, it delivered $D^{u+d}(0;\zeta_2) = -1.73(5)$.
A data-based inference yields $D^{u+d}(0) = -1.63(29)$ \cite{Burkert:2018bqq}.
Interestingly, too, within uncertainties, the CSM prediction is in $Q^2$ pointwise agreement with the empirical inference; see Ref.\,\cite[Fig.\,2]{Yao:2024ixu}.

As stressed above, nothing similar can be said about the gluon contribution to $M(Q^2)$, $\theta(Q^2)$.  Although it was long hoped that near-threshold $J/\psi$ photoproduction could provide such information, this is now known to be vain because the reaction models underlying this connection are unrealistic.
Many contemporary analyses of $J/\psi$ photoproduction provide an excellent description of available data without reference to any intrinsic property of the proton; \emph{e.g}., Refs.\,\cite{Du:2020bqj, Lee:2022ymp, JointPhysicsAnalysisCenter:2023qgg, Tang:2024pky, Sakinah:2024cza, Tang:2025qqe}.

\section{Conclusion}
The past decade has seen marked progress in strong interaction theory.
This text has only discussed properties of ground states in the pion, kaon, and nucleon channels, but looking wider, one sees a growing array of parameter-free predictions for many hadrons, including nucleon resonances \cite{Cheng:2025sdp}, not just ground states, whose properties express the full meaning of QCD.

A strong motivation for all such efforts is the need to understand how the seemingly simple QCD Lagrangian can explain the confinement of colour and emergence of the diverse and complex array of detectable hadronic states.  One may anticipate that the precise data needed to test any theoretical picture will become available during operations of existing and foreseen high-luminosity, high-energy facilities.
A validated explanation will move the scientific understanding of Nature into a new realm by, possibly, proving QCD to be the first well-defined four-dimensional quantum field theory ever considered.
If such is so, then QCD's EHM paradigm -- stressed herein -- may open doors that lead far beyond the Standard Model.

\medskip

\noindent\emph{Acknowledgments}\,---\,%
The material contained herein is the product of collaborations with many people, to all of whom we are greatly indebted.
Work supported by
National Natural Science Foundation of China, grant no.\ 12135007,
and
Helmholtz-Zentrum Dresden-Rossendorf, under the High Potential
Programme.


\end{document}